\def\year{2018}\relax
\documentclass[letterpaper]{article} 
\usepackage{aaai18}  
\usepackage{times}  
\usepackage{helvet}  
\usepackage{courier}  
\usepackage{url}  
\usepackage{graphicx}  
\usepackage{subfig}
\usepackage{balance}       
\usepackage{graphics}      
\usepackage[T1]{fontenc}   
\usepackage{txfonts}
\usepackage{mathptmx}
\usepackage{color}
\usepackage{booktabs}
\usepackage{textcomp}
\usepackage{caption}
\usepackage{subfig}
\usepackage{multirow}
\begin{document}
%
\title{AI for Trustworthiness! \\ 
Credible User Identification on Social Web for Disaster Response Agencies} 
\author{Rahul Pandey, Hemant Purohit, \and Aditya Johri\\
Geoorge Mason University\\
4400, University Dr\\
Fairfax, Virginia, 22030
\And
Jennifer Chan\\
Northwestern University \\
633, Clark St\\
Evanston, Illinois, 60208
}
\pubnote {\emph{AAAI 2018 Fall Symposium Series: AI in Government and Public Sector}}
\maketitle
\begin{abstract}
Although social media provides a vibrant platform to discuss real-world events, the quantity of information generated 
can overwhelm decision making based on that information. By better understanding who is participating in information sharing, we can more effectively filter information as the event unfolds. 
Fine-grained understanding of credible sources can even help develop a trusted network of users 
for specific events or situations. 
Given the culture of relying on trusted actors for work practices in the humanitarian and disaster response domain, 
we propose to identify potential credible users as organizational and organizational-affiliated user accounts on social media in realtime for effective information collection and dissemination. 
Therefore, we examine social media using AI and Machine Learning methods during three types of humanitarian or disaster 
events and 
 identify key actors responding to 
 social media conversations 
as \textit{organization} (business, group, or institution), \textit{organization-affiliated} (individual with an organizational affiliation), and \textit{non-affiliated} (individual without organizational affiliation) identities. 
We propose a credible user classification approach using a diverse set of social, activity, and descriptive representation features extracted from user profile metadata.  
Our extensive experiments showed a contrasting participation behavior of the user identities by their content practices, such as the use of higher  authoritative content sharing by \textit{organization} and \textit{organization-affiliated} users. This study provides a direction for designing realtime credible content analytics systems for humanitarian and disaster response agencies. 
\end{abstract}
\section{Introduction}

Social media platforms allow the public to curate and share objective and subjective information as well as interact with others online during real-world events. Increasingly, users have leveraged these affordances of social media 
across a range of socially relevant events from  
\textit{\#BlackLivesMatter} 
movements to marketing and awareness campaigns 
to natural disasters such as \textit{\#HurricaneSandy}~\cite{purohit2013emergency}. 


The easy availability of a social media platform to share information at a large scale can result in information overload, which necessitates developing a better understanding of information credibility and source types -- who is providing the information, especially where the focus is on accessing and sensemaking of high priority content (i.e. needles in the haystack). 
In the context of events relevant to disaster response agencies such as natural disasters, 
understanding the conversation among key individuals or groups within a specific trusted network is part of the work culture that may influence operational decisions of humanitarian actors~\cite{balcik2010coordination}. 
The ability to discern the 
relevant users in a trusted network of communications on social media 
can reduce information overload for sensemaking of voluminous information by pre-filtering credible information sources.  
The need 
to trust an information provider is not new within the disaster response community and there is already a culture of relying on `trusted networks' across a range of operational coordination tasks, from information sourcing to dissemination and action \cite{stephenson2006interorganizational,stoll2010interorganizational,tapia2013beyond,hiltz2014use}. Such trusted networks exist in any disaster response to an event, and can also be viewed as a response network of formally and informally related stakeholders from individuals to organizations, who pursue a common goal and purpose. 
The need for such a trusted network is critical during the time of a crisis as important decisions can be made urgently and relying on information from the trusted network reduces the perceived and real burden of assessing credibility of large volumes of information for such decision making. 
Therefore, identifying recognized actors -- whether organizational user accounts or organizational affiliated individual users -- through social media in response to a crisis event is valuable for response agencies   
%
%
and provides greater awareness for who-what-where response coordination \cite{stephenson2006interorganizational}. 
Table~\ref{tab:class-examples} shows examples of such users. 
\begin{table*}[h!]
\centering
\small
\begin{tabular}{|l|c|}
	\hline 
	\textbf{\hspace{0.1cm} Identity Class\hspace{0.1cm}} & \textbf{\hspace{0.6cm} Profile Description Text}  \\ 
	\hline 
	\textit{organization} & The official Twitter account for the American Red Cross. \\ 
	\hline 
	\textit{organization-affiliated} & CrisisMapper. Board Member of the Standby Task Force. Information Mgmt at ReliefWeb (UN OCHA) \\
	\hline 
	\textit{non-affiliated} & Calls are cool, tweets are chill, texts are alright, a Facebook message is okay \\ 
	\hline 
\end{tabular} 
\caption{Anonymized examples of user-bio descriptions of various user identity types} 
\label{tab:class-examples}
\end{table*}
For instance, during a response to a  disaster event with time-critical actions, quickly identifying user accounts of response agencies and volunteer groups as well as their affiliated users can improve identification of a pre-existing and emerging `trusted network' for response coordination 
\cite{jaeger2007community,starbird2012learning,denis2012trial,purohit2014whom}. 
This user identification process can in turn help information filtering for getting credible content written by such users for the realtime awareness of critical activities of different response stakeholders in the evolving situation ~\cite{opdyke2014building}. 
For instance, information shared by an account of an affiliated volunteer with `American Red Cross' is likely to be more trustworthy than by an individual user with no affiliation 
\cite{tapia2013beyond}. Similarly, understanding who has produced informative posts on social media 
and who else is trying to help activate the social capital around the cause of an awareness campaign (e.g., United Nations' \textit{\#HeForShe}) against societal crises like gender-based violence can provide a feedback of a `trusted network' to the non-profit organizations, who have limited resources and technologies. 

	In this study, we propose a novel AI-assisted realtime user categorization framework using diverse features of social media users 
 and specifically classify the following types of identity categories: a.) \textit{organization} - if controlled by a business, group or institution, b.) \textit{organization-affiliated} - if controlled by a person with affiliation to an organization or group, and c.) \textit{non-affiliated} - if controlled by a person without any affiliation to an organization or group. 
Inferring user identity category is a challenging task, likewise other latent attributes such as gender, ethnicity, etc. and the prior research on categorizing user identities on Twitter for event analytics from an organizational perspective has received scant attention and is underrepresented in recent studies~\cite{de2012unfolding,yan2013classifying,de2014user,park2015network,mccorriston2015organizations,oentaryo2015chalk}. 
These prior studies are limited primarily to identifying organizational user accounts. The proposed novel user identity categorization framework of diverse features and their analysis complement such studies.  
\subsection{Research Contributions.} 
 Our specific research questions are the following:
\begin{enumerate}
\item Are there distinguishable patterns of connectivity, interaction, and content practices among the relevant social media user categories to humanitarian organizations?   
\item How can we automatically identify such relevant categorical users for supporting a realtime event analytics system? 
\end{enumerate}   
\noindent The relevant user categories are 
\textit{organization} and \textit{organization-affiliated} identity users. Complementing prior research on user attribute inference on Twitter,  
we contribute via conducting the first large-scale study of diverse features for user identity classification on Twitter within the popular multiclass classification frameworks, which characterizes patterns of \textit{organizational-affiliated} and \textit{organization} identity users in contrast to \textit{non-affiliated} users. 
Specifically:  
\begin{itemize}
\item We present new empirical insights for Twitter user participation during events of interests to disaster response agencies and humanitarian organizations by conducting analyses on a variety of user, activity, and network information associated with a user identity categorization. We observed distinctive characteristics of \textit{organization}, \textit{organization-affiliated} and \textit{non-affiliated} users.  
\item We design a new ensemble-based learning framework for automatically identifying \textit{organization}, \textit{organization-affiliated}, and \textit{non-affiliated} identity users in realtime that can incorporate diverse features and is extensible.  
\item We evaluate the impacts of different user features of representation, activity, and social types extracted using only user profile metadata for the automatic identity classifier, which achieved F1-score up to 92\%. 
\end{itemize}
Our proposed approach 
can be applied for realtime analysis of participation during both slow evolving events and time-critical events such as disaster responses, given that it only relies on user profile information to derive features for automated categorization. 
It does not depend on the content of posts (also referred as \textit{tweets}) on a user's historic timeline, 
avoiding the cost of time in collecting and processing historic tweet content that hamper the capability of agencies. 

In the following, we discuss the related work and background first, followed by an analysis of metadata of users with manually labeled identities. 
After that, we describe details of the automated classification and result analysis of several experiments. Finally, we conclude with future work directions. 

\section{Background and Related Work}
\label{sect.related}

We first review existing studies that outline challenges faced by humanitarian organizations in their everyday work. We then review research on user attribute inference on social platforms, 
and with a summary of prior related works to our problem of user categorization for diverse user identities.

\subsection{Disaster and Humanitarian Context.} 
A large number of humanitarian organizations operate across the world and include both governmental and non-governmental agencies. 
Given that many organizations work across similar domains in dynamic and uncertain crisis environments, there is a need for them to coordinate their activities to avoid redundancy of efforts and to be more effective as a group. The United Nations Humanitarian Reform established the Cluster System 
in 2005 to improve coordination among both UN and non UN organizations. 
In the US context, the  National Incident Management System 
provides a structure for various agencies to coordinate while responding to incidents, threats and hazards. 
Inter-agency and inter-organizational coordination is often a significant challenge due to limitations of resources -- monetary, physical and time -- and differences in organizational culture; as a result, mutual trust is critical \cite{tapia2013beyond,denis2012trial,hiltz2014use}. There are primarily two approaches of working in such organizations, either hierarchical or distributed network of loosely coupled stakeholders \cite{stephenson2006interorganizational,balcik2010coordination}. 
The new digital age has connected these two types of organizational work environments, enabling these two cultures to increase the trusted community networks over time. 
Given the growing use of social media by such entities  \cite{dhs2014using,meier2015digital}, 
it is essential to identify relevant actors for trusted communities on social media. 
\subsection{User Attribute Inference.}
There is an extensive research on Twitter to infer user attributes, such as gender~\cite{alowibdi2013language}, age~\cite{nguyen2013old}, 
ethnicity~\cite{pennacchiotti2011machine}, etc. 
Relevant to our identity classification task, these research studies have primarily addressed the attribute inference challenge via a user classification problem and using a diverse set of features from the user's historic tweets, user profile metadata, user interactions as well as the social network structure, which guide our feature design. 
\subsection{Organization User Identification.}
The most closely related research to our proposed framework is the problem of organization account identification. Recent efforts~\cite{de2012unfolding,yan2013classifying,de2014user,park2015network,mccorriston2015organizations,oentaryo2015chalk,purohit2017classifying,karbasian2018real}  
 have classified Twitter user accounts 
with a primary focus on distinguishing  organization and personal accounts. \cite{de2012unfolding} defined and classified organizations, journalists, and ordinary individuals using features from user profile, social structure information, and historic tweets. 
\cite{yan2013classifying} defined and classified two classes of open and closed accounts, where an open account is likely to share shopping and announcement related posts, while a closed account is more likely to share personal daily-life related posts. 
\cite{de2014user} classified organization versus personal accounts using tweet content based features, while \cite{park2015network} classified group versus individual accounts using only network-based features. \cite{oentaryo2015chalk} and \cite{mccorriston2015organizations} also addressed binary classification of organization versus personal accounts, although requiring historic tweet content of users for features, while \cite{karbasian2018real} recently addressed the realtime multiclass inference of organization and other users. 
In the related crisis informatics literature, \cite{starbird2012learning} and \cite{reuter2013combining} looked into automated identification of real and virtual users in a crisis response that contain both organization or individual accounts, but non-distinguishable. A recent study \cite{purohit2017classifying}
presented a preliminary analysis of understanding \textit{organization} and \textit{organization-affiliated} users on Twitter during natural disasters, however, without any extensive analysis of distinctive feature patterns of identity classes and their importance in different classification frameworks to inform generalizability of features, across both disaster and non-disaster events.  

 The summary above suggests that existing approaches to user identity categorization are limited, given the dependence on collecting historic tweets or network structure for feature extraction, and thus, limiting feasibility for a realtime analysis as well as generalization across events. Also, not emphasizing the important class of \textit{organization-affiliated} identity users for deeper analysis also limits the fine-grained understanding of user participation patterns, especially within the disaster domain where individual affiliates may share relevant and timely information from their Twitter accounts beyond the agency accounts, due to bureaucratic communication norms. 

\section{
Approach for Classifying User Identity}
\label{sec.approach}
We propose a supervised learning approach to classify a user into a given set of identity classes---\{\textit{organization}, \textit{organization-affiliated}, \textit{non-affiliated}, \textit{none}\}, where a \textit{none} class is assigned to those users who do not belong to any of the other identity classes or cannot be determined. A user is considered to participate in an event discussion if it writes an event-related post containing a relevant hashtag or keyword. 
We first describe our data collection method in order to understand available metadata for generating diverse features for the classification framework.   
\begin{table*}[h!]
\centering
\small
\begin{tabular}{|c|c|c|c|c|c|}
	\hline 
	Event  & Duration (2016) & No. of tweets & No. of users & Sample Keywords \\ 
	\hline 		
	Louisiana  & 14 Aug. -- 30 Sep. & 2,441,805 & 833,930  & \#LouisianaFlood, Louisiana Floods, \#PrayForLouisiana,.. \\
	\hline 
	Matthew  & 06 Oct. -- 30 Nov. & 3,635,518 & 1,740,334 & \#FLPrepares, \#hurricanematthew,.. \\ 
	\hline 
GBV & 08 Nov. -- 01 Dec. & 5,334,490 & 2,214,630 & \#crimeagainstwomen, \#datedrug, lgbt harassing,.. \\ 
	\hline 
\end{tabular}
\caption{Description of diverse event datasets}
\label{table:dataset}
\end{table*}


\subsection{Event Description and Data Collection} 
We collected English language data for 3 humanitarian or disaster-specific events,
which represent diverse characteristics of social significance, time critical nature, demographics, and geographical diversity of participation, in addition to belonging to different time period of occurrence. We selected the following event topics for our study: 
\begin{enumerate}     
   \item \textit{Gender Based Violence (GBV)}. A form of global demographic event that incorporates a series of sub-events across the world. GBV is a worldwide societal crisis with diverse acts of violence \cite{russo2006gender,purohit2016gender}, including domestic assault and human trafficking. It represents a long-lasting, dynamic event that generates multitude of topical discussions on social media, and we chose the analysis period after the United States presidential elections, owing to several reports on gender issues and violence. 
    \item \textit{Hurricane Matthew (Matthew)}
.   A form of national demographic event, where a devastating hurricane affected many states in the United States, but also countries in the Caribbean, including Haiti. 
It has been classified as one of the deadliest hurricanes in the recent history.  
    \item \textit{Louisiana Floods (Louisiana)}.  
A form of local demographic event with a focus on a particular state of United States -- Louisiana. It has been classified as one of the worst natural disasters in the United States.
\end{enumerate}
\vspace{-0.4em}
We employed a keyword-based crawling approach for collecting Twitter platform data for English language tweets. 
For collecting relevant data for an event, we have a three-step process. First, we prepare a seed set of keywords relevant to an event. Second, we use the Twitter Streaming API with `filter/track' method 
to acquire a sampled stream of public tweets containing any of the seed keywords in different tweet metadata fields. 
We only considered tweets containing seed words in the text of the post. 
Third, we extract and store all the relevant metadata such as tweet text, timestamp of posting as well as authoring user's profile information such as full name, bio, and location. Our seed list of keywords for the three events included many keywords\footnote{Due to space limitation, the full keywords list, in addition to the labeled datasets, will be made publicly available with the final paper to the academic-research community.}, as shown by examples in Table~\ref{table:dataset} along with details of three events. 
\begin{table}[h!]
\centering
\small
\begin{tabular}{|c|c|c|c|}
	\hline 
	Labels & Louisiana (\%) & Matthew (\%) & GBV (\%) \\ 
	\hline 
	organization & 13.1 & 16.6 & 8.9 \\ 
	\hline 
	organization-affiliated & 7.7 & 8.8 & 7.4 \\ 
	\hline 
	non-affiliated & 68.2 & 67.5 & 76.1 \\ 
	\hline 
	none & 11.0 & 7.1 & 7.5 \\ 
	\hline 
\end{tabular}
\caption{Summary of labeled user identity classes per event}
\label{tab:labeled}
\end{table}
\subsection{Sampling for Crowdsourced Labeling} 
We selected 1500 samples from each event dataset for crowdsourced labeling. The labeling task is designed in a crowdsourcing platform, CrowdFlower\footnote{\url{https://www.crowdflower.com/}}. Annotators are required to visit the Twitter profile of a user sample and label its identity class. At least 3 judgments were taken per sample. The job instructions were refined until the annotators fully understood the labels and had consensus in judgments as per the quality control process of Crowdflower. The classes were defined as follows: 
\begin{itemize}
	\item If the user looks like an organizational or group user account, choose \textit{organization}. 
	\item If the user looks like a human with an organizational affiliation 
reported in the user profile (e.g., founder of\textellipsis, working for\textellipsis), choose \textit{organization-affiliated}. 
	\item If the user looks like a human who does not show any organizational affiliations, choose \textit{individual (non-affiliated)}.
	\item If the user looks like a bot or cannot be determined, choose \textit{none}.
\end{itemize}
The distribution of the labeled dataset is summarized in Table \ref{tab:labeled}.
\subsection{Feature Design}
\label{sec:features}
\begin{figure*}[htb]
\subfloat[Louisiana, Friends Count]{\includegraphics[trim={0 5cm 0 5cm}, width = .5\textwidth]{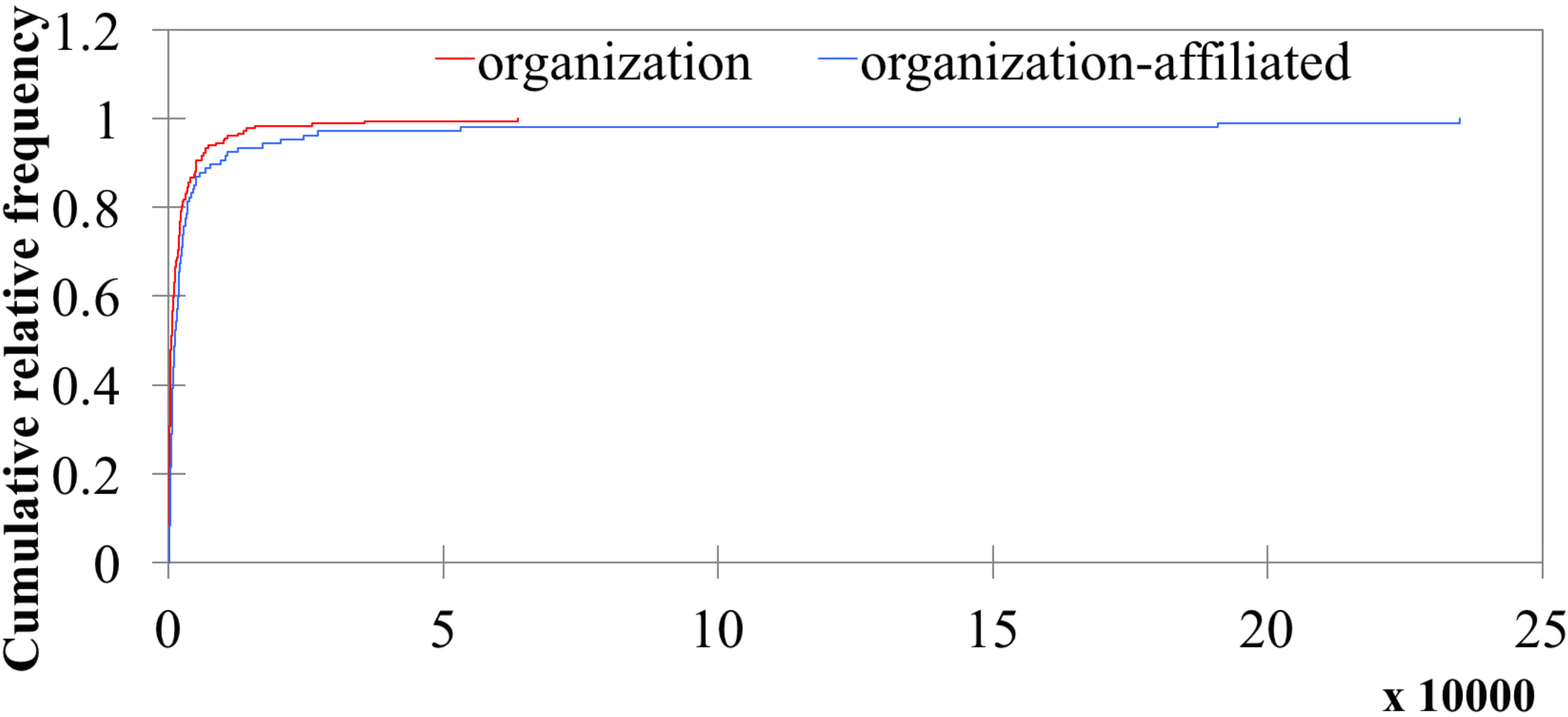}} 
\subfloat[Mattew, Favorites Count]{\includegraphics[trim={0 5cm 0 5cm}, width = .5\textwidth]{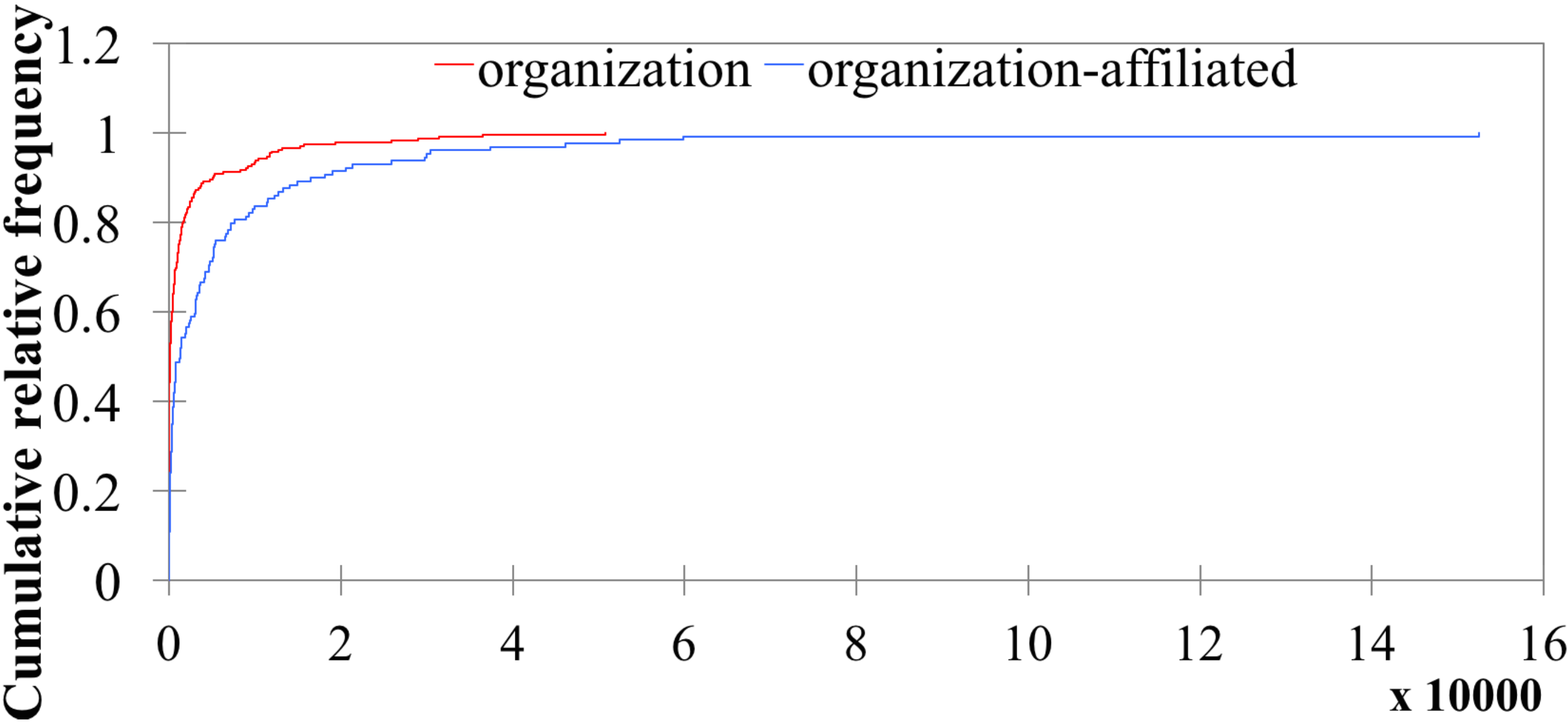}}
\\
\\
\subfloat[Louisiana, Friends Count]{\includegraphics[trim={0 5cm 0 5cm}, width = .5\textwidth]{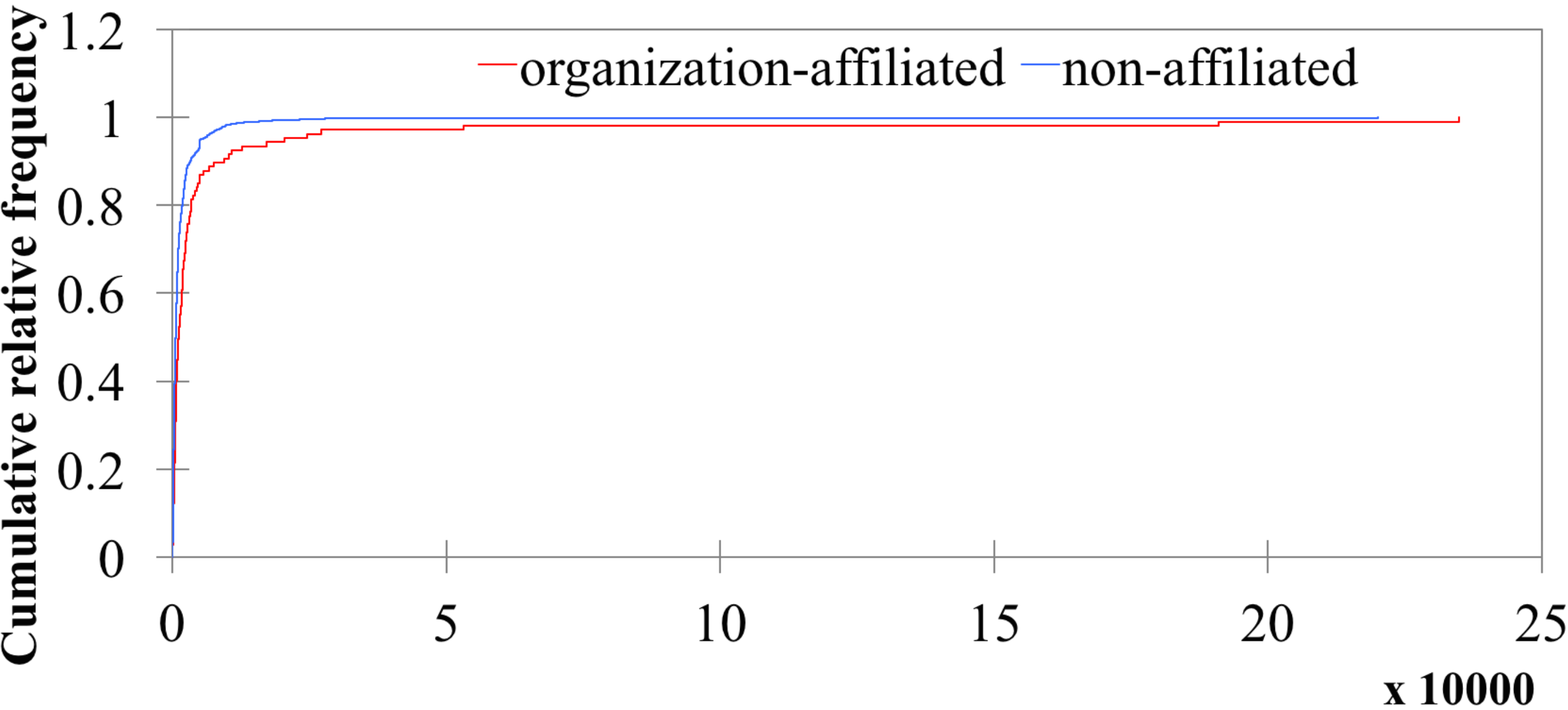}}
\subfloat[Mattew, Favorites Count]{\includegraphics[trim={0 5cm 0 5cm}, width = .5\textwidth]{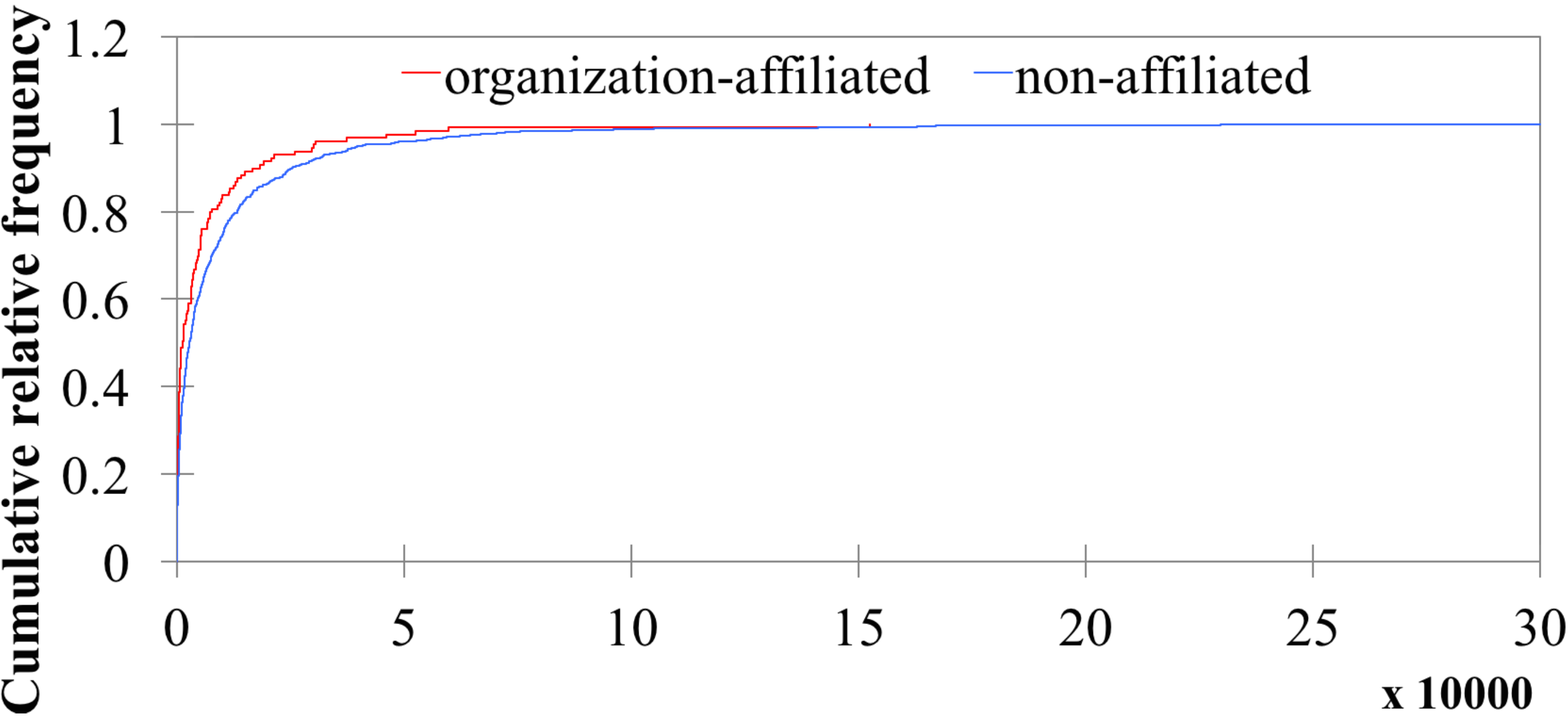}}
\caption{Illustrations of differences in the user metadata of different identity users across events, via cumulative frequency distributions. We noted distinctive characteristics across the \textit{organization}, \textit{organization-affiliated} and \textit{non-affiliated} users.} 
\label{fig:distributions}
\end{figure*}

We explored the metadata available with labeled users of different identity classes to address our research questions R1 on discriminative patterns among differing identities.  
Figure~\ref{fig:distributions} illustrates distributions of two metadata -- friend counts and favorite counts between the pairs of different identity classes to show differences (due to space limitation, we skip presenting all the plots.) To empirically investigate such differences in user metadata between identity classes, we conducted a two-sample statistical test of \textit{Kolmogorov-Smirnov (K-S) test} on the pairs of frequency distributions of labeled users from different identity classes in each event, for the metadata of follower counts, friend or followee counts, number of lists subscribed, number of statuses on the timeline, and the count of tweets favorited. 
We found statistically distinctive patterns for some metadata distributions (e.g., friends counts) 
across all the events ($p < 0.01$, rejecting null hypothesis of identical distributions), while not consistently distinctive distributions across identity pairs for a few metadata (e.g., status counts), which motivated us to propose of a mixture of diverse types of features from such metadata. 
Therefore, we design three diverse functional categories of features for developing an automated  classification framework to address our research question R2, and define novel derivative features based on the stated observations of distinctive metadata distributions -- 
 \textit{Sociability}, \textit{Favorability}, \textit{Survivability}, \textit{Activeness}, and \textit{Informality} as follows:  \\

\vspace{0.6em}
\noindent Social features: 
\vspace{-0.5em}
	\begin{itemize}
		\item \textit{Friends count} -- The number of users a user is following, to inform the user's interest-driven participation. 
		\item \textit{Followers count} -- The number of followers a user has, to inform the user's influence.  
        \item \textit{Sociability} -- Ratio of number of Friends to Followers of a user, to inform the user's social structure.
        \[ Sociability = \log (1+\frac{1+friends\_count}{1+followers\_count})\]
	\end{itemize}
\vspace{-0.1em}
Activity features: 
\vspace{-0.5em}
	\begin{itemize}
		\item \textit{Statuses count} -- The number of historic tweets of a user to inform a general degree of participation on the platform. 
		\item \textit{Favorites count} -- The number of tweets a user has favorited over time to inform a higher engagement level on the platform.  
		\item \textit{Listed count} -- The number of public lists subscribing a user, to inform user's expertise.  
        \item \textit{Favorability} -- Ratio of the number of favorites to the total number of tweets, to inform higher engagement in contrast to just posting tweets. 
        \[ Favorability = \log (1+\frac{1+favorites\_count}{1+tweet\_count})\]

		\item \textit{Survivability} -- Number of days since account creation, to inform the potential active existence on the platform over time. 
        \[ Survivability = \log (1+no.\_of\_days)\]

        \item \textit{Activeness} -- Ratio of number of tweet statuses to the number of days since account creation, to inform likelihood of a user to be active on a day on average.   
        \[ Activeness = \log (1+\frac{1+status\_count}{1+no.\_of\_days})\]

	\end{itemize}
Representation features: 
	\begin{itemize}
		\item \textit{Informality} -- a.) Number of \#, b.) Number of @ mentions, c.) number of emoticons, and d.) number of numerics used in the user's bio description.  
		\item \textit{URLs} -- Presence of URL in the user's profile metadata. 
		\item \textit{Word embeddings} -- 
        Likelihoood score of the words occurring together in user bio text, based on a trained word2vec model~\cite{mikolov2013distributed} from a corpus of user bio description text in a given event. 
 It can capture semantic coherence of users' bio representation within similar identity classes. We apply basic text preprocessing steps before training as tokenization, lowercasing, and stop-word removal.  
	\end{itemize}


\begin{table*}[h!]
\centering
\small
\begin{tabular}{|c|c|c|c|c|c|c|c|}
	\hline 
	\multirow{2}{*}{\textbf{Experiment Type}} & \multirow{2}{*}{\textbf{Feature Category}}  & \multicolumn{2}{c|}{\textbf{Louisiana}} & \multicolumn{2}{c|}{\textbf{Matthew}} & \multicolumn{2}{c|}{\textbf{GBV}}  \\ 
	\cline{3-8}
     & & Accuracy & F1-score & Accuracy & F1-score & Accuracy & F1-score \\ 
	\hline 
	\multicolumn{8}{|l|}{\textit{Single Learner}} \\
    \hline 
    baseline-1 & Social (S.) & 67.0 & 79.1 & 68.0 & 79.7 & 74.0 & 83.0 \\ 
	\hline
	baseline-2 & Activity (A.) & 71.0 & 86.0 & 71.0 & 86.1 & 76.0 & 87.8 \\ 
	\hline 
    baseline-3 & Representation (R.) & 69.0 & 78.4 & 69.0 & 78.2 & 74.0 & 82.0 \\ 
	\hline 
	Proposed & S. + A. + R. & \textbf{72.0} & \textbf{88.4} & \textbf{73.0} & \textbf{90.4} & \textbf{76.0} & \textbf{91.5} \\
	\hline 
    \multicolumn{8}{|l|}{\textit{One-vs-All Binarization}} \\
    \hline 
    baseline-1 & Social (S.) & 67.0 & 77.0 & 69.0 & 78.7 & 75.0 & 82.3 \\ 
	\hline
	baseline-2 & Activity (A.) & 71.0 & 83.2 & 71.0 & 83.6 & 76.0 & 86.1 \\ 
	\hline 
    baseline-3 & Representation (R.) & 68.0 & 77.5 & 69.0 & 77.3 & 75.0 & 80.7 \\ 
    \hline 
    Proposed & S. + A. + R. & \textbf{73.0} & \textbf{87.4} & \textbf{74.0} & \textbf{90.4} & \textbf{77.0} & \textbf{89.6} \\ 
	\hline 
    \multicolumn{8}{|l|}{\textit{One-vs-One Binarization}} \\
    \hline 
    baseline-1 & Social (S.) & 66.0 & 78.7 & 69.0 & 80.1 & 74.0 & 83.4 \\ 
	\hline
	baseline-2 & Activity (A.) & 72.0 & 84.7 & 70.0 & 85.7 & 76.0 & 88.1 \\ 
	\hline 
    baseline-3 & Representation (R.) & 68.0 & 78.0 & 69.0 & 77.9 & 74.0 & 81.8 \\
	\hline 
    Proposed & S. + A. + R. & \textbf{72.0} & \textbf{90.0} & \textbf{73.0} & \textbf{91.0} & \textbf{76.0} & \textbf{91.9} \\
    \hline
\end{tabular}
\caption{10-fold CV results for GBDT models with diverse features within the popular multiclass classification frameworks}
\label{tab:classifications}
\end{table*}

\subsection{Learning Schemes}
Our identity classification problem is a multiclass task, which is a challenging  problem and  
optimizing the performance highly depends on the feature representation, label distribution, and the problem domain. 
Therefore, researchers have studied mainly two different learning schemes to exploit such performance dependencies: a.) standalone single multiclass learner, and b.) binarization based ensemble of multiple binary (base) learners  
\cite{galar2011overview}.  
The binarization method can simplify 
learning due to exploiting distinctive characteristics for only single or two identity classes in the base learners, which is beneficial here given the observations of metadata distributions in the previous subsection, showing 
a subtle nature of difference between practices of\textit{organization} and \textit{organization-affiliated} as well as \textit{organization-affiliated} and \textit{non-affiliated} users. 
The most popular schemes of the binarization frameworks are decomposition based---One-vs-All (OVA) and One-vs-One (OVO). OVO creates a base learner for each class pair ($\textsuperscript{\emph{K}}C_2$ learners for $K$ classes). On the other hand, OVA creates a base learner for each class ($K$ learners for $K$ classes), by considering the target class instances in a positive training set, and instances of remaining classes in the negative set. 
These approaches have not been examined 
for a user identity classification. 

\section{Experiments and Results}
\label{sec.exp}
We experimented with different learning algorithms for creating identity classification models within the single as well as two binarization based multiclass classification frameworks. 
We conducted an extensive set of experiments to compare importance of diverse features within these frameworks, using well-recognized text classification algorithms ~\cite{witten2011data} --- Random Forest, Naive Bayes, Support Vector Machine, and Gradient Boosted Decision Trees (GBDT). We found the results for GBDT based classifiers to be the best; and therefore, present results for them for brevity. 
We do not have a baseline for comparison due to the lack of prior work on this classification task within such multiclass classification frameworks. Instead, to align with our research question R1 for studying discriminative characteristics of different identity users, we created three baselines for experiments corresponding to a model of each feature category (Social, Activity, and Representation) alone. We measure the classification performance using accuracy and F1-score (micro-averaged) for 10-fold cross validation (CV). 
Table~\ref{tab:classifications} presents results for the baseline models and our proposed model of all features, using the GBDT based ensemble learning algorithm, with 100 estimators.  



We observed in Table~\ref{tab:classifications} that the proposed approach of employing a mixture of diverse features (denoted by S.+A.+R.) leads to better results than the social, activity or representation category features alone, across all the multiclass classification frameworks. It is likely due to the complementary and supporting behavior of the proposed feature categories derived based on the analysis of distinctive metadata characteristics, which motivated us to exploit the combination of features. 
Next, we discuss various analyses to discover patterns of communication of the predicted users across identity classes in the large unlabeled datasets.   


\section{Analysis and Discussion}
\label{sec.discuss}
We first predicted identity class labels of unlabeled users for each event, using the One-vs-All models created from the proposed strategy of combined features, which achieved the highest accuracy. 
We do not discuss any analysis of the predicted users in the \textit{none} class due to space limitation and instead, focus on the relevant identity users. 
The distribution of relevant user identities of \textit{organization} and \textit{organization-affiliated} as well as the \textit{non-affiliated} class for the three events are: 
15.50\%, 8.85\%, and 57.03\% in Louisiana; 0.83\%, 0.10\%, and 87.69\% in Matthew; and	1.02\% 0.62\%, and 51.68\% in GBV. 
While the percentage of the relevant identity users is very small across event, albeit significant in numbers due to the scale of datasets (e.g., 73,775 \textit{organization-affiliated} in Louisiana and 22,486 in GBV) and therefore, it is important to study and detect due to their significance for helping develop a trusted community network. 
We analyzed the discriminating powers of diverse feature categories to generate such predicted class distributions first and then, discuss the social connectivity and content generation practices of the predicted users per identity. 
\subsection{\textit{Discriminative Feature Analysis}} 
We conducted $\chi^2$ \textit{test} to study importance of diverse features and ranked them. Table~\ref{tab:topFeatures} shows the set of top-5 features that are the most discriminative ones in categorizing identity classes across events. We note two important observations here. First, the top discriminative features are different 
between the two event types, 
the natural disaster versus global social crisis (GBV). It suggests different patterns of participating user identities in these events in terms of how they represent themselves in profiles (e.g., Word2Vec feature being important in GBV but not the disasters) or how they behave or engage on Twitter. This observation informs a need for further building event-specific models for the identity classification and the behavior analytics for participating user identities. Second, we note that highly discriminative features do not belong to only a specific feature category, 
supporting our proposal of combining diverse features to achieve better identity classification, as evident from the performance of the proposed combined approach in the result table (Table~\ref{tab:classifications}). We further note that as the event demographic changes from the local to the national or global crisis, `listed counts' appeared as a top feature, implying a possibility of existence of relevant identity users within global trusted networks (e.g., digital humanitarian network), manifested by their inclusion in the specific `lists' on Twitter platform. This provides a direction for future research on how we can combine prior identified trusted network with the emerging trusted networks of users as an event unfolds. 
\begin{table}[h!]
\centering
\small
\begin{tabular}{|c|c|c|c|}
	\hline 
	\textbf{Rank} & \textbf{Louisiana} & \textbf{Matthew} & \textbf{GBV} \\ 
	\hline
1 & Favorability & Listed counts & Listed counts \\ 
2 & URL usage in bio & Favorability & Mentions in bio \\ 
3 & Favorites & URL usage in bio & Word2vec Score \\
4 & Listed counts & Follower counts & Friend counts \\ 
5 & Follower counts & Favorites & Favorites \\ 
	\hline
\end{tabular}
\caption{Top features across three event datasets that belong to diverse feature categories} 
\label{tab:topFeatures}
\vskip -0.2in
\end{table}
\subsection{\textit{Social Structure Analysis}} 

Prior research has observed differences in the distributions of structural connectivity for users on Twitter~\cite{mccorriston2015organizations}. To understand the social connectivity of different user identity sets, we analyzed them by the number of friends and followers a user has. 
We observed a higher network connectivity of \textit{organizational} and \textit{organization-affiliated} users based on the mean values in terms of friend and follower outreach in comparison to the \textit{non-affiliated} users.  Interestingly, \textit{organization-affiliated} have a higher connectivity than \textit{organization} users in the natural crisis events, indicating the potential to leverage them for building trusted networks with high reachability in the network and greater likelihood to reach public during disasters. 
We further note that the higher connectivity of \textit{organization-affiliated} users for natural onset disasters than for protracted events such as GBV may also be due to the participating behavior of experts in their roles (e.g., participation by a Red Cross' 
 information officer), based upon the type of crisis and the nature of response. 


\begin{table}[h!]
\centering
\small
\begin{tabular}{|c|c|c|c|}
		\hline 
		Identity & Louisiana & Matthew & GBV \\ 
		\hline
		organization & 75\% & 78\% & 48\% \\ 
		\hline
		org-affiliated & 70\% & 66\% & 34\% \\ 
		\hline
		 non-affiliated & 69\% & 15\% & 1\% \\ 
		\hline 
\end{tabular}
\caption{Percentage of tweets containing external URLs for a user type, indicating a greater emphasis of organizational practice to attribute authoritative source of content}
\label{tab:urls}
\end{table}

\begin{table}[h!]
\centering
\small
\begin{tabular}{|c|c|c|c|}
		\hline 
		Identity & Louisiana & Matthew & GBV \\ 
		\hline
		organization & 20\% & 10\% & 20\% \\ 
		\hline
		org-affiliated & 23\% & 17\% & 13\% \\ 
		\hline
		non-affiliated & 23\% & 15\% & 16\% \\ 
		\hline 
	\end{tabular}
\captionof{table}{Percentage of tweets with `mentions' by a user type} 
\label{tab:mentions}
\vskip -0.2in
\end{table}
\subsection{\textit{Content Practices and Interaction Analysis}}
To better understand the activity and communication style patterns for the different identity users, 
we analyzed two patterns of content writing practices: 
first, inclusion of a URL in the posted content and the second, mention of other users in order to understand practices of connecting to link information sources and interactivity in the user communities. 
Table~\ref{tab:urls} shows the percentage of tweets containing URLs and Table~\ref{tab:mentions} shows the percentage of tweets containing mentions by users per identity class. 
We note two interesting observations here. First,  \textit{organization-affiliated} identity users mentions other users more than other identity users, implying they engage highly than \textit{organization} users 
under disaster events (Louisiana and Matthew), showing how they can be leveraged in times of critical events to better connect to the wider network of public. 
This behavior can also indicate the intention of actively sharing information to not only public, but also the other actors in the trusted network, usable for coordination and non-redundancy of work. 
Second, we note that both \textit{organization-affiliated} and \textit{organization} users heavily use URLs in the posts, linking to potentially authoritative sources that indicates the well-known organizational practices of evidence-based and trustworthy content sharing. 



\section{Conclusion and Future Work}

We have presented a new generic, AI-assisted user identity categorization framework for enabling trust and credibility in social media analytics via proposing three key types of user classes---\textit{organization}, \textit{organization-affiliated}, and \textit{non-affiliated}, and introducing an ensemble framework based approach. Our classifier relies on only user profile information for feature extraction that is available in realtime data stream, unlike historic tweet content based features used in the prior related work on user attribute inference for social media analytics. 

Recognizing the dynamic nature of humanitarian crisis and disasters as well as the need for a trusted network of communication and information sharing, our approach helps in filtering credible user-driven information on social media for governmental agencies and NGOs. An upstream classification system as presented here recognizes the role that trusted network play in information sharing and communication. 
This approach also recognizes the multiplicity of role that the  individuals with different user identities and groups play along the disaster/crisis continuum. It also establishes organization-affiliated users and groups as an important class with additional value. We conclude that 
filtering for a set of nails in a haystack, that being \textit{organization} and \textit{organization affiliated} users, may lead to highly relevant domain-driven content, which can be useful for humanitarian  and disaster response in complementary ways than the current literature and research to date. 
We demonstrated efficacy of diverse features in classifying the identity classes in all the well-known multiclass classification frameworks with F1-score up to 92\%.  
Our approach is useful for realtime analytics due to dependence on only user profile metadata available with streaming social media content objects, such as during disasters for realtime 
awareness of which users participate, for what purposes, and in which capacity. 

Our future work will include performing a similar study of identity classification on other social media platforms 
as well as for different types of disaster events. We will also explore \textit{non-affiliated} class user set furthermore, to develop fine-grained understanding of more user identity classes that are relevant to understanding user participation in the discussions on social media, such as the role of automated bots. 

\noindent \textbf{\textit{Acknowledgement.}} Authors thank U.S. National Science Foundation for grants IIS-1657379 and DUE-1707837. 



\balance{}

\bibliographystyle{aaai.bst}
\bibliography{aaai-fall-user-org.bib}

\end{document}